\documentclass[journal,10pt,twocolumn]{IEEEtran}
\IEEEoverridecommandlockouts
\usepackage{multirow}
\usepackage{cite}
\usepackage{amsmath,amssymb,amsfonts}

\usepackage{fancyhdr,graphicx}
\usepackage[ruled,vlined]{algorithm2e}
\usepackage{graphicx}

\usepackage{textcomp}
\usepackage{xcolor}
\usepackage{comment}
\usepackage{ulem}
\usepackage{soul, color}
\usepackage[colorlinks,allcolors=black,pdftex]{hyperref}
\usepackage{tabularx}
\usepackage{setspace}

\usepackage{etoolbox}

\newcommand{\Mname}{\textcolor{black}{Rubik}}
\newcommand{\todo}[1]{\textcolor{red}{{Yuke}: #1}}

\makeatletter
\patchcmd{\@maketitle}
  {\addvspace{0.5\baselineskip}\egroup}
  {\addvspace{-0.5\baselineskip}\egroup}
  {}
  {}
\makeatother

\begin{document}

\title{Rubik: A Hierarchical Architecture \\ for Efficient Graph Learning} 

\author{Xiaobing Chen, Yuke Wang, Xinfeng Xie, Xing Hu, \IEEEmembership{Member}, \IEEEmembership{IEEE}, Abanti Basak, Ling Liang, Mingyu Yan, Lei Deng,  \IEEEmembership{Member}, \IEEEmembership{IEEE}, Yufei Ding, Zidong Du, Yunji Chen, Yuan Xie, \IEEEmembership{Fellow}, \IEEEmembership{IEEE}\\

 \thanks{
 
 Xiaobing Chen is with State Key Laboratory of Computer Architecture, Institute of Computing Technology, Chinese Academy of Sciences, also with University of Chinese Academy of Sciences, Beijing 100190, China. 
Xing Hu, Mingyu Yan, Zidong Du, and Yunji Chen are with State Key Laboratory of Computer Architecture, Institute of Computing Technology, Chinese Academy of Sciences, Beijing 100190, China. (email:chenxiaobing@ict.ac.cn, huxing@ict.ac.cn, yanmingyu@ict.ac.cn, duzidong@ict.ac.cn,  cyj@ict.ac.cn). 
 Yuke Wang, Yufei Ding are with the Department of Computer Science, University of California, Santa Barbara, USA. 
 (email: yuke\_wang@ucsb.edu, yufeiding@cs.ucsb.edu). 
 Xinfeng Xie, Abanti Basak, Lei Deng, Ling Liang, and Yuan Xie are with the Department of Electrical and Computer Engineering, University of California, Santa Barbara, USA. 
 (email: xinfeng@ucsb.edu, abasak@umail.ucsb.edu, lingliang@ucsb.edu, leideng@ucsb.edu,  and yuanxie@ucsb.edu). Xing Hu is the corresponding author.  
 }} 

\vspace{-5pt}

\maketitle

\begin{spacing}{0.986}
\begin{abstract}
Graph convolutional network
(GCN) emerges as a promising direction to learn the inductive representation in graph data commonly used in widespread applications, such as E-commerce, social networks, and knowledge graphs. However, learning from graphs is non-trivial because of its mixed computation model involving both graph analytics and neural network computing. To this end, we decompose the GCN learning into two hierarchical paradigms: graph-level and node-level computing. Such a hierarchical paradigm facilitates the software and hardware accelerations for GCN learning.

We propose a lightweight graph reordering methodology, incorporated with a GCN accelerator architecture that equips a customized cache design to fully utilize the graph-level data reuse. 
We also propose a mapping methodology aware of data reuse and task-level parallelism to handle various graphs inputs effectively. Results show that \Mname~accelerator design improves energy efficiency by 26.3x to 1375.2x than GPU platforms across different datasets and GCN models. 

\end{abstract}

\section{Introduction}
With rich and expressive data representation, graphs demonstrate their applicability in various domains, such as E-commerce~\cite{RW-GCN,e-business_1,recommendation_gnn}, computer vision~\cite{vision_gnn1,vision_gnn2},  and molecular structures~\cite{molecular}, and \textit{etc}. To fully exploit the their value, approaches based on traditional graph analytic algorithms (\textit{e.g.}, BFS, SSSP) facilitate in-depth understanding of objects-wise relationships in graphs (\textit{e.g.}, molecule similarity in chemistry~\cite{molecular}, cells structures in bioinformatics~\cite{survey_gnn,GNN_Review}, and semantic graphs in computer vision~\cite{vision_gnn2,vision_gnn1}).  
Recently, as a rising star, extending deep learning techniques to graph analytics has gained lots of attention from both research~\cite{GraphSage,1stChebNet,FastGCN,DIFFPOOL} and industry~\cite{AliGraph,googleGraph}, largely because of their striking success on Euclidean data (\textit{e.g.}, images, videos, text, and speech). Moreover, such geometric deep learning techniques based on graph neural networks (GNNs) not only learn the inductive representations for end-to-end tasks such as classification, clustering, and recommendation, but also show much better accuracy (more than 98\%) than traditional methods~\cite{RW-GCN} (\textit{e.g.}, random walks~\cite{deepwalk}, and matrix factorization~\cite{survey_graph_embedding}). 

Among various kinds of GNNs ~\cite{GraphSage,Adaptive_Sampling}, graph convolutional network (GCN) is the most fundamental model and has been widely studied. Different GCNs can be summarized and abstracted into a uniform computing model with two stages: \textit{Aggregate} and \textit{Update}.
\textit{Aggregate} stage collects the localized node representations from the neighboring nodes.
 ~\textit{Update} stage derives the representation vector with the aggregation results. GCN distinguishes itself because of combining both neural network computing and graph computing schemes in this two stages, thus suffering from the following challenges.


\textbf{Entangled hybrid paradigms raise the difficulty of efficient computing in the uniform hardware architecture.} 
Specifically, \textit{aggregate} operation is largely based on non-Euclidean graph-level data, which is non-ordered with a diverse range of node sizes and topology. Due to the irregular memory accesses, graph-level non-Euclidean data cannot be easily handled by NN accelerators good at spatial data reuse with the statically configured vertical, horizontal, or diagonal dataflow~\cite{eyeriss, diannao}. On the other hand, \textit{update} computing consists of regular vector and matrix computations, which is computing resource hungry.
For example, GCNs features high-dimension node/edge embedding (10x -1000x\cite{KKMMN2016,GraphSage} than that of traditional graph computing) with complex NN operation (\textit{e.g.}, Multilayer Perceptron), while the traditional graph computing works with simple arithmetic operations (\textit{e.g.}, addition) on nodes with scalar values. Such computing paradigm can be hardly handled by graph accelerators with resource-intensive on-chip cache for suppress irregular accesses~\cite{graphicionado}. 
Thus, existing NN accelerator and graph accelerator designs pale in their effectiveness for handling GCN computing.

\textbf{Workload diversity and graph irregularity raise the difficulty for efficient task mapping to  utilize the hardware capability adaptively.} 
When the input graph has a larger feature dimension, the GCN computing shifts to NN computing and demands more multiply-and-Accumulator (MAC) arrays for powerful computation capability. However, when the input graph has a large number of nodes with high average degrees, the GCN computing shifts closer to graph computing that demands large on-chip buffer and data management methodology to eliminate the irregular memory access. Hence, it is essential to design an efficient task mapping methodology to bridge the gap between the diverse application demands and the uniform hardware platform. 



\vspace{10pt}
To this end, we decouple the entangled graph-level computing and node-level computing, which facilitates the software and hardware optimizations for graph learning. Such decoupled hierarchical computing paradigm is based on the following observations: 1) GCN learns both the graph-level and node-level features; 2) graph-level computing and node-level computing exhibit distinct architectural characteristics. Specifically, \textit{node-level computing} refers to intra-node computing during feature extraction and update on node-level Euclidean data with neural network techniques. Graph-level computing refers to the process of graph traversal for localized neighboring reduction  (feature reduction in Aggregate) on \textit{selected} graph-level non-Euclidean data. 

We then propose the scheduling \& mapping strategies to tackle the irregular memory access issue of the former and hardware architecture design to optimize the latter. In detail, we carry out a lightweight graph reordering on the input graphs for more graph-level data reuse potentiality. Then, we propose the programming model and tailor the neural network accelerator that incorporates a hierarchical spatial architecture with specialized cache design, to leverage the input graphs' data locality.  To bridge the gap between the diverse graph applications and uniform architectures, we propose a hierarchical mapping methodology to improve both the data reuse and task-level parallelism. 

Overall, we make the following contributions in this work:
\begin{itemize}
      \item We decouple GCN computing to two paradigms: 1) the relatively fixed and regular node-level computing, and 2) the dynamic and irregular graph-level computing. Such a decoupled computing paradigm facilitates the software and hardware optimization for GCN applications. 
      \item We propose a lightweight graph reordering method to facilitate graph-level data reuse and intermediate computation results reuse. Furthermore, we design a GCN training accelerator, \Mname, cooperated with graph reordering to support the hybrid paradigms of both node-level computing and graph-level computing.
      \item We propose the hierarchical task mapping strategies for graph-level computing and node-level computing, which comprehensively optimize both data reuse and task-level parallelism to well adapt diverse datasets with different feature sizes and graph topologies to the hardware platform. 
      \item Intensive experiments and studies show that the graph reordering and hierarchical mapping eleminates 69\% and 58\% of the off-chip memory accesses for GraphSage and GIN. 
      \Mname~outperforms GPU with 26.3x to 1375.2x of better energy efficiency.
  \end{itemize}

\section{Background} \label{sect: background}

In this section, we introduce the GCN basics, the abstract computing model, and the variants derived from GCNs. 
\subsection{GCN Basics} \label{sect: gcn basic}
The target of graph convolutional neural networks is to learn the state embedding of a graph property (node, edge, or subgraph) from the non-Euclidean input graph structure. 
Such state embeddings transform the graph features to low-dimension vectors, which are used for node, edge classification~\cite{kaspar2010graph, gibert2012graph, duran2017learning}, and graph clustering~\cite{GNN_Review}, link prediction~\cite{chen2005link, kunegis2009learning, tylenda2009towards}. In the scope of node classification tasks, we define a graph, $G = (V, E)$, where $V$ and $E$ are vertex and edge sets, respectively; each node has node feature vectors $X_v$ for $v \in V$; and each edge has edge feature vectors $X_e$ for $e \in E$.
On such a graph, GCNs learn a representation vector of a node ($h_v$), an edge ($h_e$), or the entire graph ($h_G$) with the information of the graph structure and node/edge features, so that the corresponding classification tasks can be completed based on the representations.


In terms of the computing paradigm, GCNs has two main categories: \textit{spectral GCN}~\cite{SpectralGNN1,SpectralGNN2,spectralGNN3} and \textit{spatial GCN}~\cite{GraphSage,FastGCN,DIFFPOOL,spatialGNN1, Adaptive_Sampling}. The former are derived from graph signal processing and its mathematical representation is based on eigen-decomposition of graph Laplacian matrix~\cite{1stChebNet}. 
However, spectral GCNs fall in short in several aspects: 1) The inability to perform inductive learning due to the fact that Laplacian decomposition is fixed to a specific graph; 2) The inefficiency to handle large graphs since it demands the decomposition for the entire graph adjacency matrix. 
On the other side, spatial GCNs emerge to learn the inductive representation based on the graph computing paradigm, which identifies the spatial aggregation relationships of nodes/edges. Therefore, spatial GCNs is capable to generate embeddings for unseen nodes, edges, or subgraphs. Moreover, spatial GCNs can process large graphs without compromising performance. In addition, previous works and in-depth studies~\cite{GraphSage,AliGraph,Adaptive_Sampling} also demonstrate spatial GCN as a promising direction. Base on its potential of informativeness and powerfulness, we concentrate on spatial GCN for further exploration in this work.

\subsection{GCN Computing Model}~\label{sec:computeModel}
The GCN training process consists of the following three stages: {forward propagation}, {loss calculation}, and {backpropagation}. The forward propagation calculates node feature by iteratively incorporating the impact of its neighbors, which finally outputs the status of each node comparing against the ground truth for loss computation. The backpropagation finds the impact of each state on the loss by propagating from the last layer to the input layer based on the chain rule of the gradient. It is similar as the forward propagation but in a reverse direction. 

\begin{algorithm}[h]
\small
\SetAlgoLined
\textbf{Inputs:} Graph ($V, E$); 
input features \{${X_v, \forall{v} \in V }$\};
depth $K$; 
weight matrices \{$W^k$, $\forall{k} \in K$\}; 
aggregator functions \{$AGGREGATE_k$, $\forall{k} \in K$\}; 
neighborhood function \{$N: v \rightarrow{2^V}$\} \\
\textbf{Output:} Vector representation $z_v$ for all $v\in V$
$h_v^{(0)} = X_{v} $ \\
 \For{$k = 1...K$}{
 \For{$v \in V $}{
    $a_v^{(k)} = AGGREGATE^k(\{h_u^{(k-1)}|u \in N(v)\})$
    
  $h_v^{(k)} = UPDATE^{(k)}(h_v^{(k-1)}, a_v^{(k)})$
  }
 }
 $z_v = h_v^{(k)}$
 \caption{GCN Algorithm.}
 \label{alg:gcn}
\end{algorithm}

\begin{figure}[!htbp]
    \centering
    \includegraphics[scale=0.40]{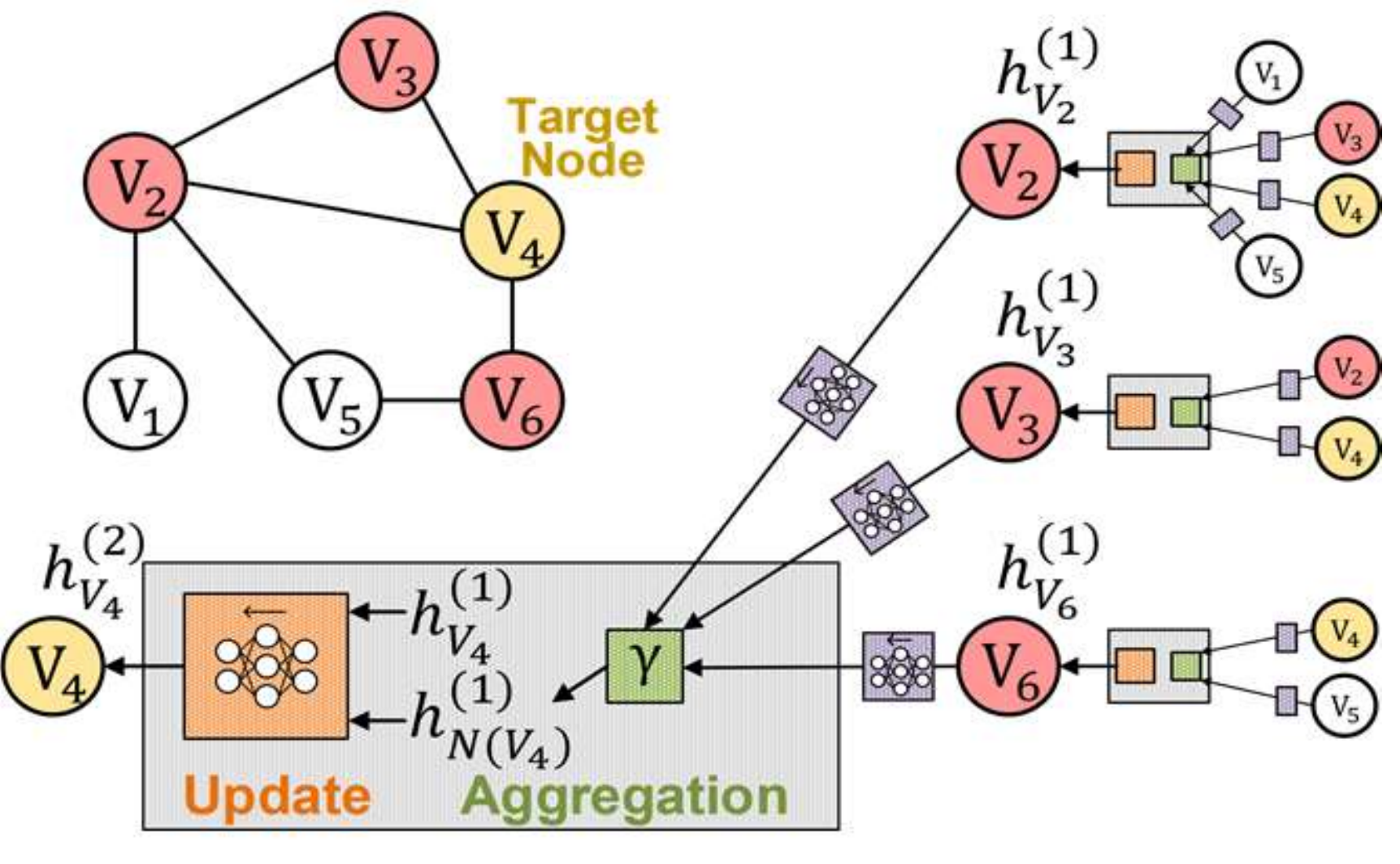}
    \caption{GCN forward propagation flow with two iterations.}
    \vspace{-1em}
    \label{fig:ff}
\end{figure}

We detail the process of the forward propagation by taking the node classification as an example. The forward propagation stage of modern GCNs works in an iterative manner, as shown at the for-loop iteration in Algorithm~\ref{alg:gcn}. Assume a node $v$ in Graph $(V, E)$ with embedding $h_v^{(0)}$ that initialized as $X_v$, and $N(v)$ refers to the set of $v$'s neighbors. $a_v^{(k)}$ and $h_v^{(k)}$ are the aggregation results and the node embedding of $v$ after the completion of the \textit{k-th} layer of a GCN. 
The computation process of GCN repeats the following two steps:  
1) Aggregate the node representation from the neighboring nodes; 2) Update the representation vector based on the aggregation results and its previous state. (some work also adopt the term of ``Combine'' instead of Update~\cite{GINConv}). 
The forward propagation process is illustrated in Figure~\ref{fig:ff} which shows the cases with two iterations. 
The backward propagation process is similar to the process shown in Figure~\ref{fig:ff} by aggregating the gradient of $a_v^{(k+1)}$ when computing the gradient of $h_v^{(k)}$.

Many variants of the functions $AGGREGATE^{(k)}(.)$ and $UPDATE^{(k)}(.)$ have been proposed to improve the prediction accuracy or to reduce the computation complexity of GCNs. 
For example, convolutional aggregators are used in graph convolutional neural networks, attention aggregators are used in graph attention neural networks~\cite{zhang2018gaan}. Gate updaters are adopted in gated graph neural networks or graph LSTM~\cite{GNN_LSTM}. 
Although there are many variants of GCN models~\cite{survey_gnn}, they can be abstracted into the uniform computing model discussed in Section \ref{sec:computeModel}.
Hence, Without loss of generality, we focus on graph convolutional neural networks in this work.

\section{Characterization in GCNs}

\subsection{Hybrid Computing Paradigms in GCN}
GCN forward propagation process is entangled with two computing paradigms: 1) the \textbf{\textit{graph-level}} computing during  node travesal and aggregating the node representations from the neighborhood in the aggregation stage; and 2) the \textbf{\textit{node-level}} computing during extracting or updating features based on deep neural network techniques. These graph-level and node-level computing paradigms demand different hardware resources. For example, neural network computing on node-level Euclidean data introduces heavy vector and matrix computation but regular memory accesses, thus dataflow optimizations can easily enlarge data reuse and eliminate the off-chip memory accesses~\cite{eyeriss}. While graph-level computing is mainly memory-bounded because of the irregular accesses in a non-Euclidean graph structure, which can be hardly handled by the data reuse strategies in neural network computing. Hence, computation and memory demands vary for different input datasets with diverse graph topology and node feature dimensions. 

\begin{figure}[!htbp]
    \centering
    \includegraphics[scale=0.40]{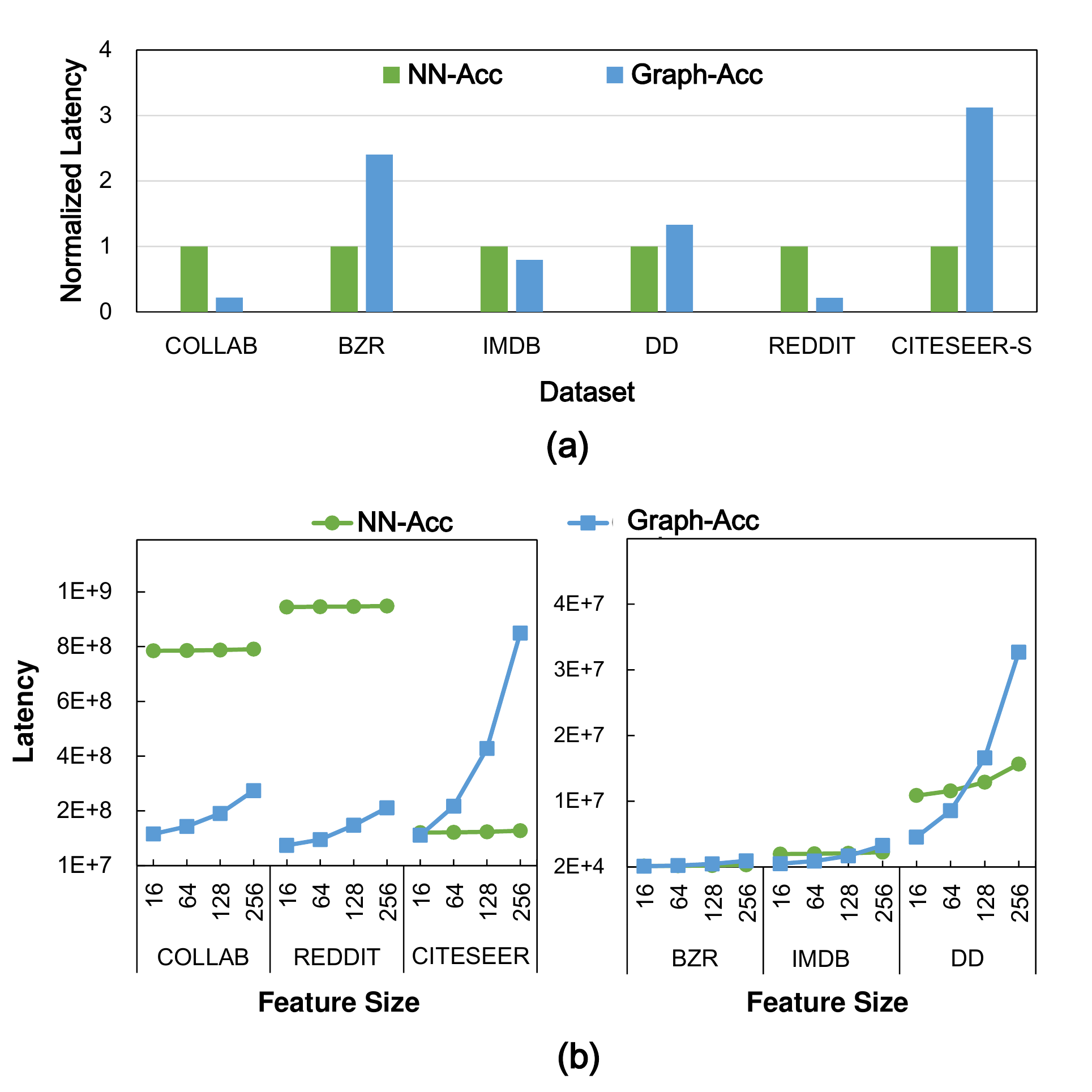}\vspace{-10pt}
    \caption{(a). Performance comparison with diverse applications. (b). Performance comparison with  different feature size.}
    \vspace{-0.6em}
   \label{fig:motivation}
\end{figure}

We further quantitatively evaluate the GCN performance of diverse input graphs with different feature sizes and degree distributions on two platforms: NN accelerator (NN-Acc) and Graph-like accelerator (Graph-Acc).  \textit{\textbf{1) NN-Acc:}}
We implement an NN accelerator similar to Eyeriss~\cite{eyeriss}, which has larger MAC arrays in every PE and has no private cache for graph traversal data buffering. The dataflow is similar to Eyeriss, which enables MACs to support efficient data reuse. The detailed configuration of the NN accelerator is shown in Table~\ref{table:config}. \textit{\textbf{2) Graph-Acc:}} 
We tailor the graph accelerator to execute the graph convolutional neural networks. The Graph accelerator closely resembles a prior Graph accelerator~\cite{graphicionado}, which is equipped with a large on-chip buffer and the processing array to deal with the matrix-vector multiplication. The detailed configuration is shown in Table~\ref{table:config}. 
We evaluate six GCN datasets on GIN (detailed configurations are in Section~\ref{sec:setup}) and the results are shown in Figure~\ref{fig:motivation}. We have the following observations:

 1) \emph{Computing input graphs with lower degree shifts to NN computing mode and favors more computing resources}. For example, \textit{BZR}, \textit{DD}, and \textit{Citeseer-S} have an average degree of 1.1, 2.5, 3.6, NN accelerator performs better than the Graph accelerators.
 
 2) \emph{Optimization for non-Euclidean graph-level data reuse plays a much more important role for training input graphs with a larger average degree}. For example, \textit{COLLAB}, \textit{IMDB-BINARY}, and \textit{REDDIT} have an average degree of 32.8, 4.8, and 492. Thus, Graph accelerator performs better than NN accelerator. 
 
3) \emph{NN accelerator is extremely under-utilized because of the memory inefficiency for most of the GCN models.}
Taking the \textit{REDDIT} in Figure~\ref{fig:motivation}(b) as an example, the execution latency of the NN accelerator stays still even the output dimension scales from 16 to 256, which indicates that the computation capability is under-utilized and NN accelerator is heavily memory-bounded which largely incurred by the graph irregularity.  
 
 In summary, GCNs favor \textit{NN-Acc} with powerful computation capabilities and optimizations for spatial data reuse when the input graph has a high feature dimension, while GCNs appreciate \textit{Graph-Acc}  with larger on-chip memory when input graphs exhibit high irregularity and complex topologies. Thus, there are two important issues to be addressed for designing GCN acceleration architectures: 1) how to optimize the memory access efficiency of graph-level data; and 2) how to design efficient and feasible architectures for input graphs
with diverse graph scales and feature dimension sizes when algorithms constantly evolve.

\subsection{Opportunities in Graph-level Data Reuse}

We observe that there are two different types of data reuse opportunities in GCNs: node-level (Euclidean) and graph-level (non-Euclidean) data reuse. 
Taking the illustrative case in Figure~\ref{fig:reuse}(b) as an example, during the update stage, feature vectors of $node_6$ are fed in the neural networks as input. Such neural network computing for node-level data has been well studied in the previous work~\cite{eyeriss}. Thus the spatial architectures that exploit high compute parallelism using direct communication between processing elements (PEs)
can be used to optimize the data reuse in either vertical, horizontal, or diagonal directions~\cite{eyeriss}. 

During the graph feature computation in the aggregation reduction stage, the irregular memory access cannot be efficiently handled by the Euclidean dataflow methodologies that exploit high spatial locality through using direct communication between processing elements in either vertical, horizontal, or diagonal directions. 
However, because of the intrinsic graph feature in the real-world graphs, such as ``community'' structure that some nodes share neighbors or have denser connections to a group of nodes, thus offers two types of graph-level data reuse opportunities: \textit{graph-level feature data reuse (G-D)} and \textit{graph-level computation results reuse (G-C)}. 

\subsubsection{\textbf{Graph-level feature data reuse}} The node feature data can be potentially reused during graph traversal in the aggregation computation.  
As shown in Figure~\ref{fig:reuse}, when computing neighbor aggregation for $node_6$, feature data of $node_4$, $node_5$, and $node_8$ will be accessed. When computing neighbor aggregation for $node_2$, $node_4$ and $node_5$ will be accessed. Hence, the feature data of $node_4$ and $node_5$ will be repetitively reused if we traverse the graph for aggregate computing with the order of $node_2$ and $node_6$. Such data reuse of node feature data during graph traversal is referred to as graph-level feature data reuse. The reuse distance is determined by the graph topologies and traversal order. 

\subsubsection{\textbf{Graph-level aggregation computation reuse }} 
\begin{figure}[h]
    \centering
    \includegraphics[scale=0.22]{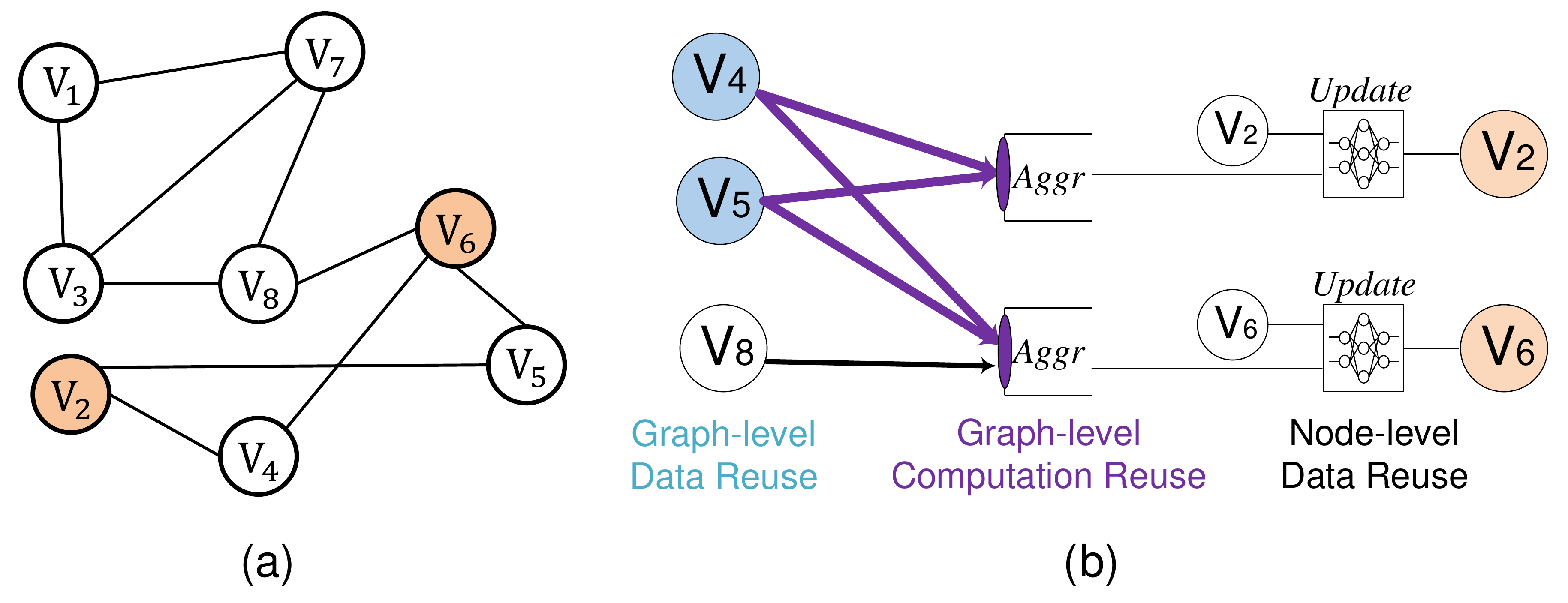}
    \caption{(a) An example of input graph; (b) Data reuse schemes: graph-level data reuse, graph-level intermediate computation reuse, and node-level data reuse.}
    \label{fig:reuse}
\end{figure}

The intermediate aggregation results can be potentially reused because of the shared neighbor sets in the ``community" structure of graphs and the order-invariant feature of aggregation operators.  The aggregation reduction operations are commonly based on \textit{sum}, \textit{average}, or \textit{min/max}. The computing order doesn't affect the final result. Hence the intermediate computation results of shared neighbor sets can be reused. 
For example, the $node_2$ and $node_6$ have the shared neighbor sets: $node_4$ and $node_5$. The intermediate results of aggregating $node_4$ and $node_5$ can be reused when computing the $node_2$ and $node_6$, as illustrated in Figure~\ref{fig:reuse}(b). Benefits of computation reuse come from two folds: 1) eliminating the useless redundant computing of feature vectors; 2) alleviating the memory burden and data thrashing during redundant computation of node feature vectors.  

In summary, significant volume of graph-level data locality hide during the irregular graph traversal.
With the limited on-chip memory resources, graph scheduling strategies are important to reduce the data reuse distance for more efficient memory accesses.


\begin{figure}[h]
\centering  
\includegraphics[scale=0.30]{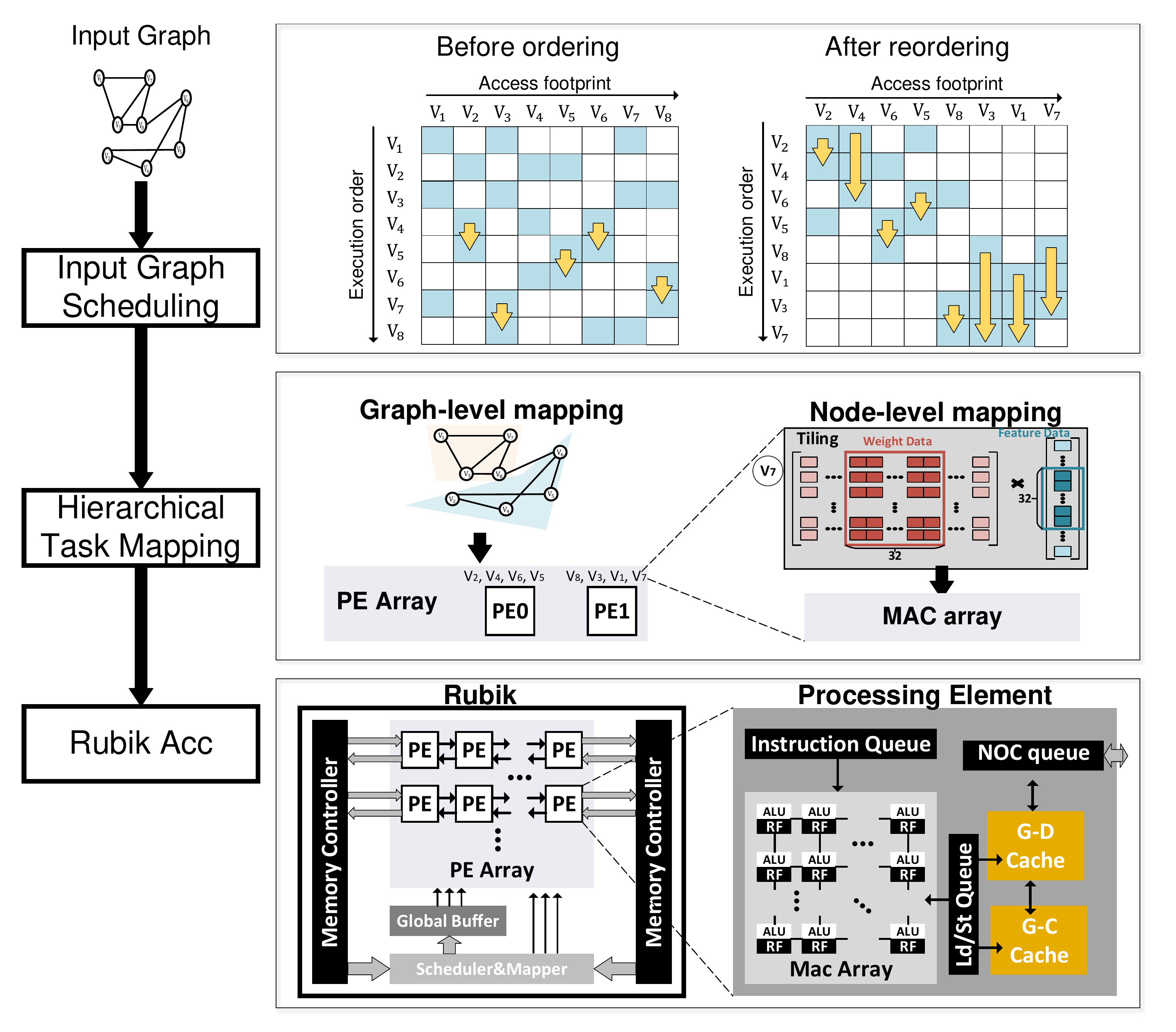}
\caption{Design overview of Rubik: 1) Input graph reordering that groups the nodes with more shared neighbors together to reduce reuse distance; (2) Hierarchical task mapping; (3) The Rubik architecture.}
 \label{fig:DesignOverview}
\vspace{-1em}
\end{figure}

\begin{figure*}[!th] \small
    \centering
    \includegraphics[scale=0.40]{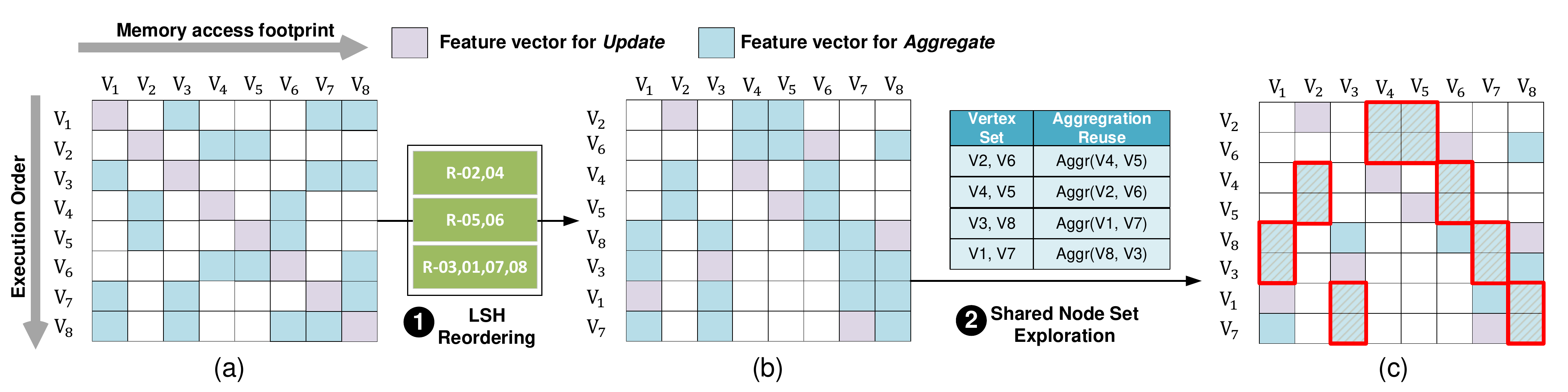}
    \vspace{-1.5em}
    \caption{Input graph reordering: (a) Index order before ordering; (b) LSH-based row reordering; (c) Shared-Set Exploration.}
    \label{fig: Row-Column Reordering.}
\end{figure*}

\vspace{0.2em}
\section{\Mname~Design}
In observations of the challenges and opportunities of GCN applications, we propose \Mname~to fully utilize both the graph\&node level data locality and computation parallelism. The key design concept is to decouple the entangled non-Euclidean computing and Euclidean computing, propose software-based methodology to optimize the former and hardware architecture design to optimize the latter. 

As shown in Figure~\ref{fig:DesignOverview}, \Mname~mainly consists of three parts: 1) the input scheduling methodology that utilizes graph reordering to determine the traversal order for smaller reuse distance of graph-level feature data and computation results; 2) the mapping methodology to allocate the tasks to processing elements for both computation parallelism and data reuse. 3) the hardware architecture design that cooperates with scheduling and mapping methodology to leverage both the global-level and node-level data locality. We enhance neural network accelerators in a lightweight way, so that minimum effort is needed to tailor the neural network accelerator for efficient GNN computing.

\subsection{Input Graph Ordering}
In this section, we introduce a lightweight graph reordering methodology which improves the graph-level data locality. 
In this work, the graph reordering happens at the pre-processing stage for only once. Such ordering can be integrated in the graph pre-processing in GNN algorithms that adopt the graph topology features for more efficient training~\cite{clusterGCN}. More discussion about the overhead and feasibility is in Section~\ref{sec:discuss}.   

The goal of reordering is to group the nodes with more shared neighbors together to improve the graph-level data reuse when conducting aggregation reduction operations. The intrinsic reason that the reordering method can provide better temporal reuse is based on the fact that real-world graphs exhibit a ``community structure" ~\cite{Graphcommunity}, which means that some nodes share neighbors or have a closer relationship to each other. Therefore, by grouping them together, the data locality during execution will be significantly improved. Note that graph reordering does not change the graph structure but only affects the execution order of the graph. We develop the graph reordering algorithm by synergistic \textit{Locality-Sensitive Hashing} and \textit{Row-Column Ordering}.  

\vspace{0.2em}

\noindent\textit{\textbf{1) LSH-based Graph Reordering}}



\emph{Locality-Sensitive Hashing (LSH)} is an algorithmic technique, being widely used to solve the approximate or exact nearest neighbor problem in high dimension space~\cite{LSH1,LSH2,LSH3}. It groups similar items into the same ``buckets" with high probability. The basic concept based on random projection: for every input vector \textit{x}, the hash function is calculated by projecting this vector \textit{x} to several random vectors. With a series of random vectors, LSH maps an input vector to a bit vector (buckets). Input vectors with smaller distances have a higher probability to result in the same cluster with the same bit vector.

\emph{Reordering Flow:}
We leverage the LSH technique to cluster the nodes with more shared neighbors. Every row in the adjacency matrix of the graph is a vector that represents the neighbor connections for this vertex. Taking these vectors of rows in the adjacency matrix, LSH hashing groups the rows into several clusters. Taking the input graph in Figure~\ref{fig:reuse}(a) as an example, the processing flow is illustrated in Figure~\ref{fig: Row-Column Reordering.}. \textit{Row-02} and \textit{Row-04} are grouped in the cluster because they share most of the neighbors and have similar row vectors. Similarly, \textit{Row-05} and \textit{Row-08} are grouped in a cluster. \textit{Row-03}, \textit{Row-01}, \textit{Row-07}, and \textit{Row-08} are grouped in a cluster. Thus, after the row transformation in step 1, we have the transformed graph with the nodes assigned to the same buckets being placed continuously, as illustrated in Figure~\ref{fig: Row-Column Reordering.}(b).  In this way, the reuse distance of the node feature vectors are reduced.  


\vspace{0.2em}
\noindent\textit{\textbf{2) Shared Node Set Exploration}}

Based on the reordered graph, we explore the reuse potentiality of the intermediate aggregation computation results. The basic idea is to find the shared node sets in the window of neighboring rows. A simple example is illustrated in Figure~\ref{fig: Row-Column Reordering.}(c), where $V_2$ ($node_2$) and $V_6$ share the neighbor set of $V_4$ and $V_5$. Therefore, the intermediate results of $V_4$ and $V_5$ can be reused for $V_2$ and  $V_6$ aggregation computation. Similarly, the intermediate 
computation result of  $V_1$ and  $V_7$ can be reused for  $V_3$ and  $V_8$. 

Considering it is too time-consuming to obtain shared node sets that maximize the potential computation results reuse, we adopt an alternative heuristic by limiting the search window instead, \textit{i.e}, finding the shared node set only in adjacent nodes in the execution order.  For instance, ($V_2$, $V_6$), ($V_4$, $V_5$), ($V_3$, $V_8$), ($V_1$, $V_7$) in the simplified case illustrated in Figure~\ref{fig: Row-Column Reordering.}(b).


\subsection{Hardware Accelerator}
We tailor the neural network accelerator  to fully utilize the graph-level data locality. Specifically, \Mname~ accelerator supports both spatial and temporal data flow for regular (node-level) and irregular (graph-level) computing, enhanced with both G-D cache and G-C cache for graph-level data reuse and computation reuse. 

The architectural design of \Mname~ is demonstrated in Figure~\ref{fig:DesignOverview}. \Mname~ is mainly comprised of the following basic components: \textit{processing element (PE) array}, \textit{on-chip memory hierarchy}, and \textit{control logic}. 

\vspace{0.2em}
\noindent\textit{\textbf{1) PE Array}}

The overall PE array is hierarchically organized based on multiple PEs constituted by MAC arrays. The graph-level computing tasks are dispatched to PE array and node-level computing tasks are scheduling inside the MAC array, for the ease of programming and optimization for utilization. Multiple PEs connected with the 2D-mesh network on chip (NoC) interconnections. The leftmost and rightmost PEs in the NoC mesh communicate with the memory controller directly. The other central PEs get the read and send write requests through the 2D-mesh NoC. All the traffic between PEs and memory goes through the NoC network in a first-come-first-serve manner with the one-way routing strategy. There are two memory controllers in \Mname~at the left and right side of PE arrays. The access location of a memory request is determined by the memory address. Once the access location is determined, the memory request is transferred through the NoC at either left-horizontal or right-horizontal directions.

The detailed design of PE is shown in Figure~\ref{fig:DesignOverview}(e), which consists of the \textit{instruction queue}, \textit{load-store queue (LSQ)}, \textit{NoC queue}, \textit{multiply and add accumulator (MAC) array}, and two \textit{private caches} (G-C and G-D cache) for data reuse of graph-level non-Euclidean data. 
Instruction Queue buffers the micro-instructions including three major categories: \textit{load}, \textit{store}, and \textit{computation}. The entire GCN training process can be translating to hardware primitives according to the input graphs. The detailed programming model and hardware primitives are in Section~\ref{sec:program}. 
The micro-instructions can be generated by the driver and prefetched to the instruction queue with the streaming strategy for good access efficiency. LSQ buffers the load and store requests for accessing the feature extraction data, aggregation data, and update data. G-D and G-C caches store the graph-level feature data and computation results. 


\vspace{0.2em}
\noindent\textit{\textbf{2) On-chip Memory Hierarchy}}

Hierarchically, the on-chip memory is comprised of \textit{global buffer} for PE array, \textit{private G-D and G-C cache} in every PE, and \textit{register files (RFs)} in every MAC. Global buffer exploits data reuse between PEs, such as the weight matrices. Except for the weight metrics, all other store requests are write-through and sent back to the memory controller directly without on-chip buffering. MAC register files are similar to that in NN accelerators which exploit all types of data movement within the computations of one node, including the convolutional reuse and filter reuse during node-level computing.  

Private G-D and G-C caches exploit graph-level data locality in a temporal manner, by buffering the feature vectors and intermediate aggregation results of graph-level non-Euclidean data inside every PE. Tasks in different PEs do not have non-Euclidean data reuse nor any data dependency, in order to improve task-level parallelism with out any cache conflict. It is important to well adapt GNN applications with diverse graph scales and feature dimension sizes to hardware accelerators with careful consideration about task parallelism and data reuse efficiency. The detailed mapping methodology is introduced in Section~\ref{sec:mapping}. 

The working flow is as follows: during the calculations of aggregation operations, PE first tries to search feature vectors of neighbors in G-D cache. If it is not a hit,  PE gets the feature vector data from off-chip memory, and then stores the feature vector data of neighbors in G-D cache. If the computation reuse optimization option is initiated, PE searches the G-C for the intermediate aggregation results with the tag of node index ids. If it is a hit, the results will be obtained directly for the following computation, which eliminates the redundant computing. Otherwise, PE will search G-D again for feature vectors of neighbors individually. For the ease of implementation and reduce the storage overhead of tag bits, the reuse of intermediate aggregation results is at the granularity of two nodes. Both G-C and G-D cache adopt the LRU (least recently used) replacement strategy since graph ordering stage already optimizes the reuse distances. 

\begin{figure*}[!th] \small
    \centering
    \includegraphics[scale=0.35]{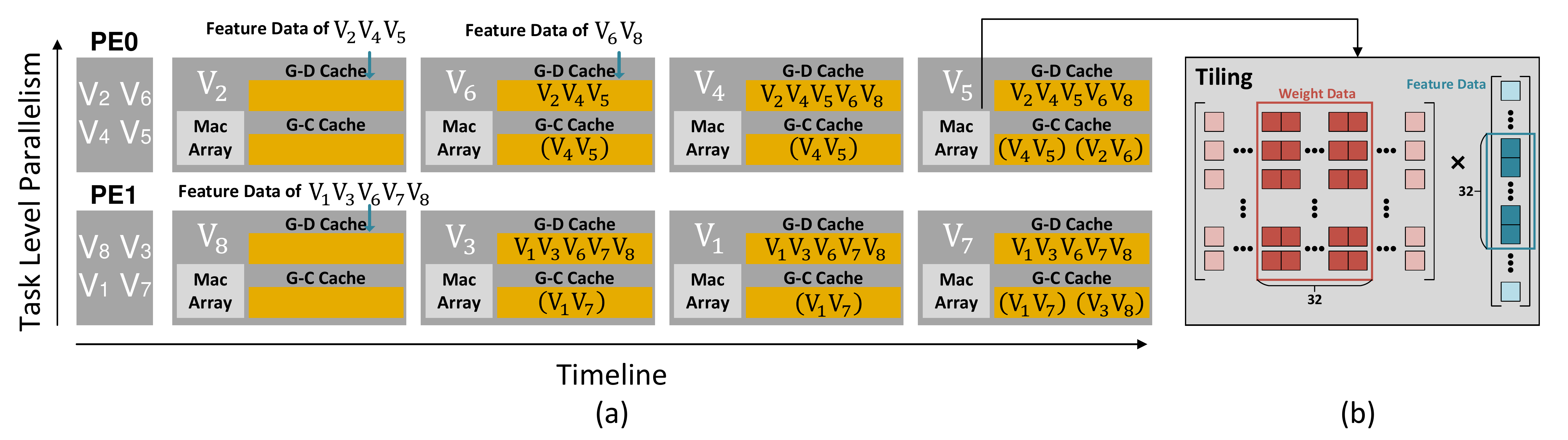}
    \vspace{-1em}
    \caption{Hierarchical task mapping: (a) Graph-level mapping (Node sets allocation in PEs); (b) Node-level mapping (Intra-node task tiling in MAC array).}
    \label{fig:taskmapping}
\end{figure*}

\subsection{Programming Model and Hardware Primitives}~\label{sec:program}
To generally support diverse GCN algorithms, we adopt a vertex-centric programming model, since most graph neural networks are based on this model~\cite{GraphSage,clusterGCN,FastGCN,Adaptive_Sampling}. Based on the vertex-programming model, we provide the following hardware primitives to support the execution of GCNs in  Algorithm~\ref{alg:gcn}: \textit{load-f}, \textit{load-i}, \textit{comp}, and \textit{store}. 
The first two primitives, \textit{load-f} and \textit{load-i}, are used to load and aggregate the feature vector of single node and the intermediate aggregation result of two nodes. 
The third primitive, \textit{comp}, is used to invoke the computation of feature extraction and update function, which is usually composed of matrix-vector multiplication and some element-wise computation instructions.
After computing the feature vector of a node $v$ for the $k$-th layer ($h_v^{(k)}$), the \textit{store} primitive is used to flush the result of computation into memory so that it is visible to other PEs in the iteration of $(k+1)$-th layer.

Using vertex-centric programming models has no need to worry about the data conflict issue in edge-centric programming, but is confronted with synchronization issues during execution. Such overhead is introduced when the node update operation is blocked due to waiting for neighbors to be aggregated. Thus, we propose a graph reordering method and intelligent mapping to alleviate irregular memory access effect and the corresponding synchronization overhead, while retaining the task-level parallelism. 
The reordering and mapping stage generate two inputs to the hardware accelerator. The first input is the task assignment with the ordered vertex ID. Each PE is assigned with a set of vertices to compute.
The second input is the indicator for the reuse of the intermediate aggregation results, which generates the \textit{load-i} instructions. 
The hardware accelerator executes these hardware primitives generated by the reordering and mapping stages, which exploit the locality of feature vectors and the computation reuse of partial intermediate results.

\vspace{0.2em}
\subsection{Mapping Methodology}~\label{sec:mapping}
With the input reordered graph, we map the tasks onto the \Mname~accelerator in a hierarchical manner. Specifically, task mapping first partitions the input graph, and decides the node set allocations to every processing element, which is referred to as \textit{graph-level mapping}. Then the intra-node computations are organized into MAC arrays, which is referred to as \textit{node-level mapping}.

\begin{figure*}[h]
\centering  
\includegraphics[scale=0.55]{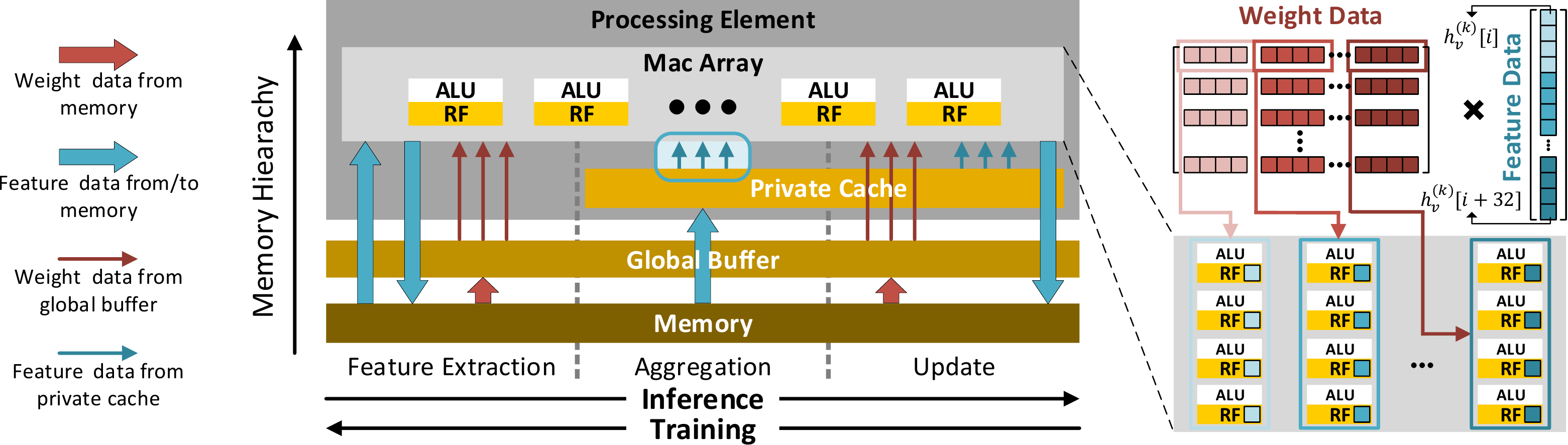}
\caption{Dataflow in Rubik: an example of weight data, feature data, and intermediate aggregation results reuse across four layers of memory hierarchy.}
 \label{fig:dataflow}
\end{figure*}

\vspace{0.2em}
\noindent\textit{\textbf{1) Graph-level mapping:}}
The mapping strategy of allocating vertices to PEs considers both data reuse and task-level parallelism. 
After graph reordering, the nodes in the traversal sequence have a similar set of neighbors, which enables both the input data reuse and intermediate computational result reuse. Hence, we allocate the consecutive nodes in a window of reordered traversal sequence in one PE, while every individual PE computes a different window for task parallelism. 

Taking the Graph in Figure~\ref{fig: Row-Column Reordering.} for instance, the execution order after ordering is $V_2$, $V_6$, $V_4$, $V_5$, $V_8$, $V_3$, $V_1$, and $V_7$. With the window size of 4, $V_2$, $V_6$, $V_4$, $V_5$ are allocated in $PE_0$, while $V_8$, $V_3$, $V_1$, and $V_7$ are allocated in $PE_1$. 
Such a process is illustrated in Figure~\ref{fig:taskmapping} (a). 
In $PE_0$, $V_2$, $V_4$, $V_5$, and $V_6$ will be executed sequentially. During computing the aggregation operations for $V_2$, feature data of $V_4$
and $V_5$ are obtained from off-chip memory and buffered in G-D cache. Since there is an indicator of shared node sets of $(V_4, V_5)$, the intermediate aggregation results of them will be stored in G-D cache for further reuse. When computing the aggregation operations for $V_6$, feature data of $V_2$, $V_8$, and intermediate results of $(V_4, V_5)$ are needed. $V_2$ and $(V_4, V_5)$ are hit in G-D cache and G-C cache respectively, therefore we only need to get the feature data of $V_6$ and $V_8$. When computing the aggregation and update operations of $V_4$ and $V_6$, all the feature data of neighbors are in cache and no off-chip memory traffic is introduced during computation. Such an example shows that the graph-level tasking mapping based on the reordered graph improves the temporal reuse locality for the vertices in the same PE.

\noindent\textit{\textbf{2) Node-level mapping:}}
For the feature vector computation inside nodes (feature extraction and update), we tile the vector-matrix multiplication onto the MAC arrays for a better data reuse. Such mapping and tiling techniques have been well studied in previous work~\cite{eyeriss,diannao}. We adopt a similar methodology, as shown in Figure~\ref{fig:mapping}(b). The matrix-vector multiplication is partitioned to several blocks according to the computation capability of MAC arrays. 

In summary, such a hierarchical task mapping method decouples the irregular graph mapping and regular node mapping for better data reuse and computing parallelism.

\subsection{Dataflow in different Computation Stages}
In this subsection, we introduce the computing and data reuse process of the whole forward propagation. The backpropagation is similar but in a reverse way. As introduced in Section~\ref{sect: background}, the whole processing pipeline of the forward propagation is comprised of 
\textit{aggregation reduction} and \textit{update}. The detailed forward propagation computation and dataflow are shown in Figure~\ref{fig:dataflow}.

Overall, the data reuse in \Mname~can be generally classified into two categories: the reuse of \textit{graph-level data} and \textit{node-level data}. For the node-level computation, such as feature extraction and update stages, the feature map data and weight matrix are reused in MAC arrays. For the graph-level computation on the node set for aggregation, the feature data is stored in the private cache of every PE for temporal reuse. 

 Feature extraction for nodes is initiated at the beginning of every iteration. During this process, the feature data of nodes are streaming in and streaming out to memory systems. Weight data is stored onto the global buffer and reused for the feature extraction of every node. 

\vspace{0.2em}
\emph{Aggregation.}
After the completion of feature extraction, \Mname~conducts aggregation reduction for every node, by loading and computing the feature data of its neighbors. The feature data is buffered in the private (G-D and G-C) cache of PE. Along with the aggregation for the nodes in the input graph, there is temporal reuse of feature map data in G-D cache and intermediate aggregation results in G-C cache. Such temporal data reuse reduces off-chip memory accesses and the Section~\ref{sec:mapping_results} discussed the effect with a quantitative analysis. 

\emph{Update.}
With the aggregation results of a node as input, the update operation is carried on by  calculating the aggregation results and the previous state of this node. During such a regular computing process, the weight data and feature data are reused in MAC arrays and the global buffer. The final result of the updated feature data will be written through to off-chip memory directly. 

\section{Experimental Results}
In this section, we first introduce the experimental setup and analyze the performance impact of graph reordering and mapping methodologies. Then we compare the performance and energy of ~\Mname~ to NN accelerator, GPU, and CPU 
Finally, we analyze the impact of embedding size and graph degree on performance and show that ~\Mname~ can well adapt diverse applications on the hardware platform. 
\begin{figure*}[h] \small
    \centering
    \includegraphics[scale=0.45]{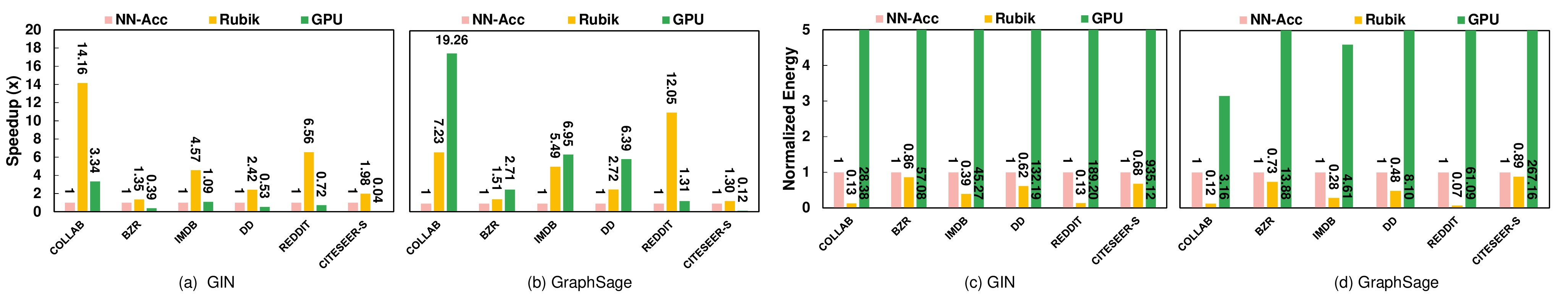}
    \caption{Speedup and energy comparison for NN-like accelerator, Rubik, and GPU.}
    \label{fig:perf}
\end{figure*}
\begin{figure*}[h] \small
    \centering
    \includegraphics[scale=0.43]{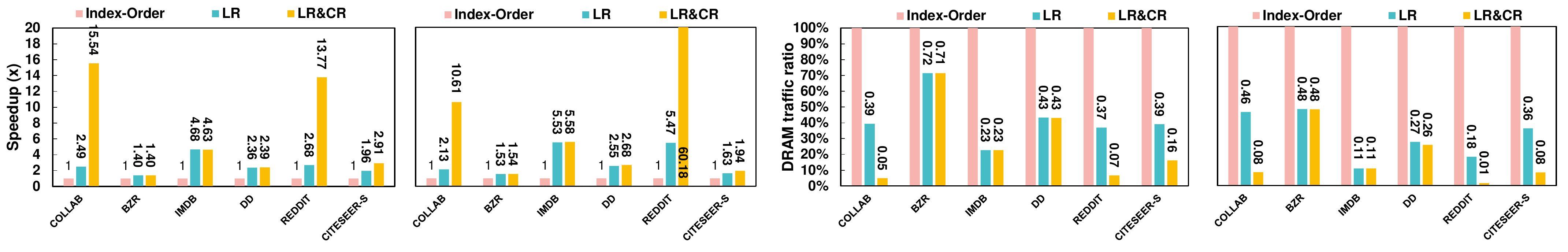}
    \caption{Speedup and off-chip memory traffic reduction under different graph scheduling\&mapping strategies.}
    \vspace{-1em}
    \label{fig:mapping}
\end{figure*}

\subsection{Experimental Setup}\label{sec:setup}
\vspace{0.2em}
\noindent\textbf{GCN Datasets.}
Our Graph accelerator evaluation covers a wide spectrum of mainstream graph datasets, including benchmark datasets for graph kernels~\cite{KKMMN2016}, and datasets commonly used by previous studies~\cite{GraphSage,clusterGCN} in related domains. 
Details of these datasets are listed in Table~\ref{table: Graph Datasets}.  We also build a synthetic benchmark of the citeseer~\cite{nr}, named Citeseer-S,  
which has 227,320 vertices with the dimension of 3,703. 
Such a relatively large graph with a high dimension is built to test the hardware capability. 
\begin{table}[h] \small
\centering
\vspace{-1em}
\caption{Graph Datasets}\label{table: Graph Datasets}
\vspace{-8pt}
\small
\scalebox{0.9}{
\begin{tabular}{| c | c  c  c  c  c |}
\hline
\textbf{Dataset} & \textbf{\#G} & \textbf{Avg.\#V} & \textbf{Avg.\#E} & \textbf{D} & \textbf{\#Class} \\
\hline
COLLAB      & 5,000	& 74.49	    & 2,457.78   & 492 & 3   \\
BZR         & 405 	& 35.75	    & 38.36     & 53  & 2  \\
IMDB-BINARY & 1,000	& 19.77	    & 96.53     & 136 & 2  \\
DD          & 1,178	& 284.32    & 715.66    & 89  & 2  \\
\hline
\hline
\textbf{Dataset} & \textbf{\#G} & \textbf{\#V} & \textbf{\#E} & \textbf{D} & \textbf{\#Class} \\
\hline
CITESEER-S     & 1           &	227,320      &   814,134      &   3,703    & 41     \\
REDDIT        & 1           &	232,965      &	114,615,892   &   602    & 6     \\
\hline
\end{tabular}}

\end{table}

\vspace{0.2em}
\noindent\textbf{GCN Models.}
In this work, we mainly test on two commonly-used graph convolutional neural network models: \textbf{GIN}~\cite{GINConv} and \textbf{GraphSage}~\cite{GraphSage}. We use the default configuration in broadly-used GCN library (Pytorch Geometric (PyG)~\cite{PyTorch_Geometric}), where GraphSage has 2 sageConv layers with hidden dimension = 256,  GIN has 5 sageConv layers and 2 linear layers with hidden dimension = 128. 

\vspace{0.2em}



\noindent\textbf{Hardware Configurations.}
\subsubsection{\textbf{\Mname}}
We implement a cycle-accurate simulator to evaluate the total execution latency (cycles), while the accelerator is working conservatively at 500Mhz, as simulated in Section~\ref{sec:exp_hw}. This simulator models the modules in the architecture design, including PE, NoC, on-chip buffer, private cache, and \textit{etc}, as introduced in Section 4. The configuration of ~\Mname~ is shown in Table~\ref{table:config}. 
\subsubsection{\textbf{GPU}}
In addition to accelerators, we also evaluate the GCN performance on NVIDIA Quadro P6000 GPU (3840 CUDA cores, 12TFLOPs peak performance, 24GB GDDR5X memory, 432GB/s peak bandwidth). The GCN implementations are based on PyG~\cite{PyTorch_Geometric}. The GPU performance is estimated 
by NVProf~\cite{nvprof}, which eliminates the memory copy time and system stack overhead.

\begin{table}[]
\caption{Hardware Platform Configurations}\label{table:config}
\vspace{-5pt}
\scalebox{0.71}{
\begin{tabular}{|c|c|c|l|c|c|c|c|}
\hline
\multicolumn{2}{|c|}{} & \multicolumn{2}{c|}{\textbf{NN-Acc}} & \textbf{Graph-Acc} & \textbf{Rubik} & \textbf{GPU} \\ \hline
\multirow{2}{*}{\textbf{Comp}} & PE Array & \multicolumn{2}{c|}{8x8 PEs} & 8x8 & 8x8 & \multirow{2}{*}{3840 Cores} \\ \cline{2-6}
 & MAC Array & \multicolumn{2}{c|}{16x16 MACs} & 1x4 & 4x8 &  \\ \hline
\multirow{4}{*}{\textbf{Mem}} & Mem BW & \multicolumn{4}{c|}{32GB/s} & 432GB/s \\ \cline{2-7} 
 & Global buffer & \multicolumn{2}{c|}{2 MB} & 4 MB & 2 MB & L2: 3MB \\ \cline{2-7} 
 & Private Cache & \multicolumn{2}{c|}{--} & 256KB/PE & 128KB/PE & L1:48KB/SM \\ \cline{2-7} 
 & \multicolumn{1}{l|}{RegisterFile} & \multicolumn{2}{c|}{16KB/PE} & \multicolumn{1}{l|}{256B/PE} & \multicolumn{1}{l|}{2KB/PE} & \multicolumn{1}{l|}{RF: 48K/SM} \\ \hline
\end{tabular}}
\vspace{-1.5em}
\end{table}

\subsection{Scheduling Optimization}~\label{sec:mapping_results}
\Mname~incorporates both the hardware accelerator design and mapping methodology based on the reordered graph. In this section, we first analyze the impact of graph reordering which aims to improve the data reuse of non-Euclidean data. Specifically, we compare the following three strategies on \Mname~platform: 1) \textit{Index-order:} compute with the index order of nodes; 2)  \textit{LSH-Reordering (LR)}: compute the nodes in the reordered order after 
LSH-based graph reordering; 3) \textit{Reordering\&Computation results Reuse (LR\&CR)}:  reuse the intermediate aggregation computation results in the G-C cache, with the reordered input graphs. 

\noindent\textbf{Performance Comparison.} We compare the speedup of the latter two strategies over the first one, as shown in Figure~\ref{fig:mapping}(a) and (b). We make the following observations: 1) Reordered graph generally improves the performance with the speedup of about 3.14x and 2.59x for GraphSage and GIN, across the datasets with different degree distributions and feature dimension sizes. 2) For input graphs with larger degrees, reusing the graph-level intermediate computation results \textit{(LR\&CR)} brings significant speedup. As shown in Figure~\ref{fig:mapping}, COLLAB has an average degree of 32 and it achieves 15.5x speedup by reusing the aggregation results during GIN training.         
\vspace{0.2em}

\noindent\textbf{Off-chip Memory Traffic Reduction.}
We further analyze the off-chip memory access reduction with dataflow optimization.  
The off-chip memory access volume of these three strategies is shown in Figure~\ref{fig:mapping}(c) and (d). Generally, compared to index-order execution, \textit{LR} graph reordering reduces 69\% and 58\% of the off-chip memory access traffics for GraphSage and GIN. For the large sparse graphs with a large average degree, such as COLLAB and Reddit, the intermediate aggregation reuse \textit{(LR\&CR)} eliminates more than 90\% of memory accesses in the further step. 
These results consistantly show that optimization for non-Euclidean data significantly reduces the memory traffic and improves the memory access efficiency. 


\subsection{Speedup}
 We compare the performance and energy efficiency of NN accelerator (baseline), ~\Mname, CPU, and GPU, with the detailed configurations shown in Table~\ref{table:config}. For the fair of comparison, all these architectures take in the same re-ordered graphs.

\noindent\textbf{Performance.}
We evaluate the execution latency of training the entire graph for one epoch and compare it with the baselines, as shown in Figure~\ref{fig:perf}(a) and (b). Overall, \Mname~shows speedups of 
1.35x to 14.16x compared to
NN accelerator when running GIN model. Meanwhile, \Mname~achieves 
1.30x to 12.05x of speedup when running GraphSage.

We further compare \Mname~with the GPU platform and provide the following observations.

1) \emph{Larger graphs with high dimension size and node volumes are more performance-sensitive to the data reuse optimizations}. When training GraphSage models, \Mname~achieves 9.18x and 10.76x of speedup for Reddit and Citeseer-S with a large graph scale. While GPU outperforms \Mname~when training small graphs, such as COLLAB, BZR, IMDB, and DD. The key reason is that their memory footprint is too small and most feature data and weight data can be held in the on-chip memory hierarchy thus training GCNs becomes computing-bound. For larger graphs, the feature data of nodes cannot be held in the on-chip hierarchy. Additionally, in GCNs, most of the operations are based on matrix-vector multiplication, which has a much larger mem/compute ratio than that of matrix-matrix multiplications. Thus the data reuse optimization plays a more important role for larger graphs. Consistently, \Mname~achieves a larger speedup compared to GPU on Reddit and Citeseer-S.

2) \emph{Deeper GCN models are more performance-sensitive to the data reuse optimizations}. GIN model, which has deeper layers (5 Sageconv layers and 2 linear layers) than that in GraphSage (2 SageConv layers), \Mname~achieves the speedup of 3.42x to 4.52x compared to the GPU platform even on small graphs (COLLAB, BZR, IMDB, and DD). Overall, \Mname~achieves the speedup of 3.42x to 46.7x of speedup across the various datasets when training GIN models. 

\vspace{2pt}
\subsection{Hardware overhead}~\label{sec:exp_hw}
 We compare the performance and energy efficiency of NN accelerator, ~\Mname, and GPU, with the detailed configurations shown in Table~\ref{table:config}. 
For the power and area evaluation of NN and \Mname~accelerators, we break down the circuit model estimation to the compute logic, memory array, and hierarchical wires. We adopt Design Compiler under 45nm technology for RTL synthesis of MAC array and control logic, Micron Power Calculators for SRAM and DRAM estimation, McPAT~\cite{mcpat} for  the NoC interconnection area and power estimation. We conservatively run the accelerator at 500Mhz, which comfortably satisfies the timing restraints. GPU power is sampled by \textit{nvidia-smi}, which is the tool suite provided by NVIDIA CUDA driver. 

\noindent\textbf{Energy Consumption.}
In addition to performance comparison, we compare the energy consumption of \Mname,  NN accelerator, and GPU. 
Energy consumption is calculated by multiplying the average power and the execution time. 
Compared to GPU, \Mname~improves energy efficiency by 26.3x to 1375.2x across different datasets and GCN models.  Compared to NN-like accelerators, \Mname~improves energy efficiency by 1.47x to 7.92x for GIN and 1.13x to 8.20x for GraphSage. 
For graph-like accelerators, \Mname~improves energy efficiency by 1.60x to 1.87x for GIN and 1.69x to 2.52x for GraphSage. 
Such a relatively smaller energy consumption gap from the graph-like accelerator than the gap from the NN-like accelerator is caused by the large proportion of energy consumption on the on-chip cache and DRAM memory access.

\vspace{2pt}
\noindent\textbf{Area.} We further evaluate the area of head of \Mname~, which mainly consists of the following components: \textit{computation logic}, \textit{on-chip buffer and queues}, \textit{hierarchical interconnection}, and \textit{control logic}. The computation units comprise of the MAC arrays. The on-chip buffer and queues include the LSQ, instruction queue, G-D Cache, G-C Cache, global buffer, and register file, as described in Table~\ref{table:config}. In summary, under the technology process of 45nm, \Mname~has an area of 36.86 $mm^2$.

\vspace{-0.3em}

\vspace{0.2em}
\section{Discussion}~\label{sec:discuss}
\begin{figure}[h]
    \centering
    \includegraphics[scale=0.35]{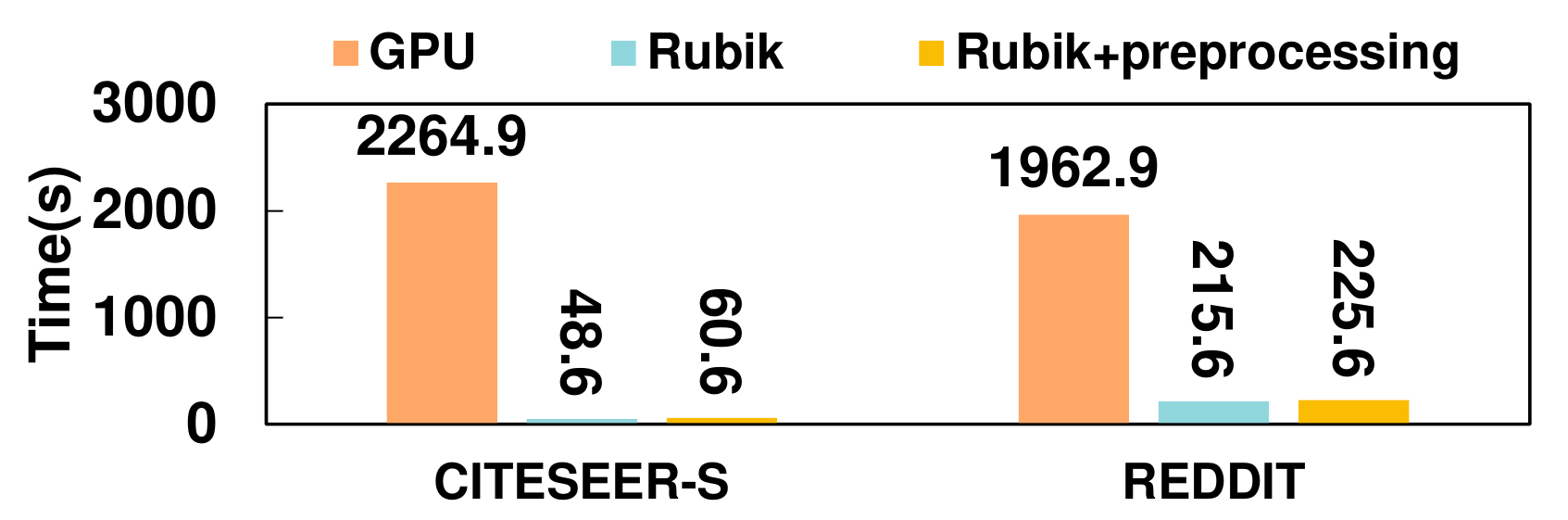}
    \caption{Preprocessing overhead.}
    \vspace{-1em}
    \label{fig:preprocessing}
\end{figure}

\vspace{0.1em}
\textbf{Graph-Reordering Overhead.}
In this work, the graph reordering is happening in the pre-processing stage for only once. It is based on row and column transformation according to the LSH clustering results. LSH clustering is lightweight and friendly for hardware parallelization.  For the Reddit dataset with 232,965 nodes, the graph reordering only requires several seconds to complete. We compare the performance between GPU and \Mname~ with/without preprocessing overhead under the training scenario with 100 epochs, as shown in Figure~\ref{fig:preprocessing}. Without preprocessing overhead, \Mname~ achieves 46.7x and 9.06x of speedup on Citeseer and Reddit. With preprocessing overhead, \Mname~ still achieves 37.4x and 8.66x speedup compared to GPU. 

In addition, such an LSH-based technique can be extended to support on-line graph reordering for batching and sampling techniques. The LSH-clustering has the time complexity of $O(n*nz*|H|)$, where $|H|$ is the number of the hashing functions, and $nz$ is the average non-zero elements in the adjacency matrix. Supporting the on-line graph reordering will be our future work. 
\vspace{-1em}


\textbf{Batching and Sampling Influence.}
Batching and sampling strategies are proposed to train the graph model to alleviate the memory and computation burden for training the entire graph data in one epoch and improve the convergence speed as well~\cite{GraphSage,clusterGCN}. The state-of-art algorithm work~\cite{clusterGCN} observes that the training node sets with more edges are very important for improving the convergence rate of the GCN models during sampling or batching. Our reordering methodology greatly helps to target the node sets with large dense connections, thus enabling a more efficient batching and sampling method. Additionally, the reordered graph remains useful even for random batching or sampling because the order for temporal data reuse stays still in the subgraphs. 

\section{Related Work}
\vspace{-0.2em}
\textbf{Graph acceleration.}
In observing that the graph applications exhibit the high cache miss rates and under-utilization of memory bandwidth, abundant works have been proposed to accelerate graph analytics applications. They can be classified as the following categories: \textbf{1) Graph Preprocessing}: In order to improve the data access efficiency, it is necessary to preprocess graph data that adapts the graph structure to the hardware accelerators. For example, graph layout reorganization, graph ordering~\cite{graphOrdering_IISWC18}, and graph partitioning~\cite{graphpartition1}. Our work incorporates the graph ordering techniques to improve the data reuse of non-Euclidean dataflow during GCN training. \textbf{2) Hardware acceleration}: Customized architectures have been proposed to accelerate graph applications. Previous work designs hardware modules to implement the gather, apply, scatter phases in graph computing~\cite{graph_accelerator_ozturk,graphicionado}.  Graphicionado adopts large on-chip eDRAM for storage of the graph data to eliminate the random accesses, and another work~\cite{graph_accelerator_ozturk} designs a dedicated cache hierarchy for different graph data. However, such an on-chip design cannot efficiently handle the spatial data reuse inside the NN-based  computation. Additionally, the computing units in graph accelerators are too lightweight for the NN-based computation of GCN applications. 

\vspace{0.2em}
\textbf{DNN Accelerators.}
Academia and industry have proposed various architectures for the general acceleration of DNNs~\cite{dnnAcc_ms,diannao,eyeriss,DNNAcc_Minerva}, which can be classified as the temporal architectures and spatial architectures. The spatial accelerators are based on dataflow processing, where the processing element or ALUs form a processing datapath for directly communicating with each other. Many advanced dataflow optimization strategies are proposed, such as input stationery, weight stationery, and row stationery, \textit{etc}. Such dataflow designs eliminate the overhead of loading or storing data from and into memory hierarchy. However, the dataflow optimizations are only applicable to Euclidean data processing with regular data reuse directions or datapaths. For the irregular graph data, there is no uniform data reuse datapath. Therefore, our work propose a memory hierarchy design to support both of these two dataflows to improve data access efficiency.   

\vspace{0.2em}
\textbf{GNN Accelerators.}
In observing the challenges of GNN computing, some pioneering work have been proposed to accelerate the GCN inference. Yan \textit{et al}. ~\cite{hyGCN} and Auten \textit{et al}.~\cite{dacGNN} propose the accelerator design for GNN networks with pure hardware design. Yan's work proposes the hardware methodology, window sliding and window shrinking, to improve memory efficiency. However, as we demonstrated, processing index-order input graphs ignore the global-level data locality. Our work decouples the hierarchical paradigms and leverage two schemes of graph-level data locality for feature data reuse and intermediate aggregation result reuse, achieving better performance speedup.
\vspace{-10pt}
\vspace{0.1em}
\section{Conclusion}
The graph convolutional network (GCN) is a promising approach to learn the inductive representation of graphs from many application domains.
To meet the demands of this new learning method mixing the computation of graph analytics and neural network, we propose the geometric learning accelerator based on spatial architectures for graph neural network models, \Mname, and enhance memory hierarchy design to support the data reuse of both the Euclidean and non-Euclidean data. 
We further develop a lightweight graph reordering strategy to improve the temporal reuse of non-Euclidean data and eliminate workload.
Finally, we evaluate \Mname~accelerator design and compare it with the existing architectural design of the NN accelerator and graph accelerator on representative GCN models and datasets. 
Evaluation results demonstrate that \Mname~together with our mapping method achieves significant speedup and better energy efficiency compared with prior designs.
\end{spacing}

\bibliography{refs}

\end{document}